\documentclass{ws-procs9x6}

\def\nabstar#1{\nabla\kern-0.5pt\smash{\raise 4.5pt\hbox{$\ast$}}
               \kern-4.5pt_{#1}}

\def\drvstar#1{\partial\kern-0.5pt\smash{\raise 4.5pt\hbox{$\ast$}}
               \kern-5.0pt_{#1}}

\def\newline{\relax\ifhmode\null\hfil\break\else\nonhmodeerr@\newline\fi}
\def\frac#1#2{{#1\over#2}}
\def\text#1{{\hbox{\rm #1}}}

\newcommand{\beq}{\begin{equation}}
\newcommand{\eeq}{\end{equation}}
\newcommand{\bea}{\begin{eqnarray}}
\newcommand{\eea}{\end{eqnarray}}
\def\Id{ \mbox{1\hspace{-1.2mm}I} }

\def\BA{\begin{eqnarray}}
\def\EA{\end{eqnarray}}
\def\BAN{\begin{eqnarray*}}
\def\EAN{\end{eqnarray*}}

\def\tr{\mbox{tr}}

\def\gm5{\gamma_5}

\usepackage{graphicx}
\begin{document}

\title{Signal of $ \Theta^+ $ in quenched lattice QCD
       with exact chiral symmetry\footnote{\uppercase{T}his work  
was supported in part by the
\uppercase{N}ational \uppercase{S}cience \uppercase{C}ouncil, 
\uppercase{ROC}.}}


\author{Ting-Wai Chiu, Tung-Han Hsieh}

\address{Physics Department and National Center for Theoretical Sciences
at Taipei \\ 
National Taiwan University, Taipei, Taiwan 106, Taiwan}

\maketitle

\abstracts{
We investigate the mass spectrum of the pentaquark baryon ($ udud \bar s $) 
in quenched lattice QCD with exact chiral symmetry.  
Using 3 different interpolating operators,  
we measure their $ 3 \times 3 $ correlation matrix  
and obtain the eigenvalues $ A^{\pm} (t) $ with $ \pm $ parity,   
for 100 gauge configurations generated with Wilson gauge 
action at $ \beta = 6.1 $ on the $ 20^3 \times 40 $ lattice. 
For the lowest-lying $ J^P = 1/2^- $ state, 
its effective mass is almost identical to 
that of the KN s-wave,  
while for the lowest-lying $ J^P = 1/2^+ $ state,
its effective mass is smaller than that of the 
KN p-wave, especially for the regime $ m_u < m_s $.
By chiral extrapolation (linear in $m_\pi^2$) to $ m_\pi = 135 $ MeV, 
we obatin the masses of the lowest-lying states: 
$ m(1/2^-) = 1424(57) $ MeV and $ m(1/2^+) = 1562(121) $ MeV, 
in agreement with the masses of $ m_K + m_N \simeq 1430 $ MeV and   
$ \Theta^+(1540) $ respectively. 
}

\section{Introduction}

The recent experimental observation of the exotic baryon $ \Theta^+(1540) $ 
(with the quantum numbers of $ K^+ n $)
by LEPS collaboration\cite{Nakano:2003qx} 
at Spring-8 and the subsequent confirmation  
from some experimental groups has become one of the most interesting topics 
in hadron physics. 
The remarkable features of $ \Theta^+(1540) $ are its strangeness $ S=+1 $,
and its exceptionally narrow decay width ($ < 15 $ MeV) even though it is 
$ \sim 100 $ MeV above the KN threshold.
Its strangeness $ S = +1 $ immediately  
implies that it cannot be an ordinary baryon composed 
of three quarks. Its minimal quark content is $ u d u d \bar s $. 
Nevertheless, there are quite a number of experiments\cite{Hicks:2004ge}
which so far have {\it not} observed $ \Theta^+(1540)$ or any pentaquarks. 
This casts some doubts about the existence of $ \Theta^+(1540) $. 

Historically, the experimental search for $ \Theta^+(1540) $ was motivated 
by the predictions of the chiral-soliton model\cite{Diakonov:1997mm}, 
an outgrowth of the Skyrme model.
Even though the chiral solition model 
seems to provide very close predictions for the mass and the width of 
$ \Theta^+(1540) $, obviously, it cannot reproduce all aspects of QCD, 
the fundamental theory of strong interactions.      
Now {\it the central theoretical question is whether the spectrum of QCD 
possesses $ \Theta^+ $ with the correct quantum numbers, mass, 
and decay width.} 

At present, the most viable approach to solve QCD nonperturbatively
from the first principles is lattice QCD. 
Explicitly, one needs to construct an 
interpolating operator which has a significant overlap with 
the pentaquark baryon states. Then one computes the time-correlation
function of this interpolating operator, and from which to extract
the masses of its even and odd parity states respectively. 
However, since any $ (udud\bar s) $ operator must couple to 
hadronic states with the 
same quantum numbers, it is necessary to disentangle 
the lowest lying pentaquark states from the $ KN $ scattering states, 
as well as the excited pentaquark states. 

To this end, we adopt the 
variational method\cite{Michael:1985ne,Luscher:1990ck},
and use three different interpolating operators 
of ($ udud \bar s $) to compute their $ 3 \times 3 $ correlation 
matrix, and from its eigenvalues to extract the masses of the 
even and odd parity states.   
These three interpolating operators (with $ I=0 $) are
\cite{Mathur:2004jr,Zhu:2003ba,Csikor:2003ng,Jaffe:2003sg,Sugiyama:2003zk,Sasaki:2003gi,Chiu:2004gg}:
\bea
\label{eq:O1}
{(O_1)}_{x\alpha} &=&
  [{\bf u}^T C \gamma_5 {\bf d}]_{xc} \
\{ \bar{\bf s}_{x \beta e} (\gamma_5)_{\beta\eta} {\bf u}_{x \eta e}
 (\gamma_5 {\bf d})_{x \alpha c}
\nonumber\\ && \hspace{10mm} 
- \bar{\bf s}_{x \beta e} (\gamma_5)_{\beta\eta} {\bf d}_{x \eta e} 
(\gamma_5 {\bf u})_{x \alpha c} \}
\\
\label{eq:O2}
{(O_2)}_{x\alpha} &=&
  [{\bf u}^T C \gamma_5 {\bf d}]_{xc} \
\{ \bar{\bf s}_{x \beta e} (\gamma_5)_{\beta\eta} {\bf u}_{x \eta c}
   (\gamma_5 {\bf d})_{x \alpha e}
\nonumber\\ && \hspace{10mm} 
- \bar{\bf s}_{x \beta e} (\gamma_5)_{\beta\eta} {\bf d}_{x \eta c} 
 (\gamma_5 {\bf u})_{x \alpha e} \}   \\
\label{eq:O3}
{(O_3)}_{x\alpha} &=& \epsilon_{cde}
  [{\bf u}^T C \gamma_5 {\bf d}]_{xc} \
  [{\bf u}^T C {\bf d}]_{xd} \ (C \bar {\bf s}^T)_{x\alpha e}
\eea
where ${\bf u}$, ${\bf d}$ and ${\bf s}$ denote the quark fields;
$ \epsilon_{abc} $ is the completely antisymmetric tensor;
$ x $, $ \{ a, b, c \} $ and $ \{ \alpha, \beta, \eta \} $ 
denote the lattice site, color, and Dirac indices respectively;
and $ C $ is the charge conjugation operator. 
Here the diquark operator is defined as 
\bea
[{\bf u}^T \Gamma {\bf d} ]_{xa} \equiv \epsilon_{abc} ( 
 {\bf u}_{x\alpha b} \Gamma_{\alpha\beta} {\bf d}_{x\beta c}
-{\bf d}_{x\alpha b} \Gamma_{\alpha\beta} {\bf u}_{x\beta c} )
\eea 
where $ \Gamma_{\alpha\beta} = -\Gamma_{\beta\alpha} $.
Thus the diquark transforms like a spin singlet ($1_s$), 
color anti-triplet ($ \bar 3_c $), 
and flavor anti-triplet ($ \bar 3_f $). For $ \Gamma = C \gamma_5 $, it
transforms as a scalar, while for $ \Gamma = C $, it transforms like 
a pseudoscalar. 
Note that $ O_1 $, $ O_2 $, and $ O_3 $ all transform like an even 
operator under parity.

In the Jaffe-Wilzcek model\cite{Jaffe:2003sg}, each pair of $ [u d] $ 
form a diquark. 
Then the pentaquark baryon $ \Theta([ud][ud] \bar s) $ emerges as the 
color singlet in 
%
 $  ( \bar 3_c \times \bar 3_c) \times \bar 3_c 
   = 1_c + 8_c + 8_c + \overline{10_c} $, 
and a member (with $S=+1$ and $I=0$) of the flavor anti-decuplet in 
%
 $ \bar 3_f \times \bar 3_f \times \bar 3_f   
  = 1_f + 8_f + 8_f + \overline{10}_f $. 
Now, if one attempts to construct a local interpolating operator
for $[ud][ud]\bar s $, then these two identical diquarks must be
chosen to transform differently (i.e., one scalar and one pseudoscalar),  
otherwise $ \epsilon_{abc} [ud]_{xb} [ud]_{xc} \bar s_{x\alpha a} $
is identically zero since diquarks are bosons.
Thus, when the orbital angular momentum of this
scalar-pseudoscalar-antifermion system is zero
(i.e., the lowest lying state), its parity is even rather than odd.
Alternatively, if these two diquarks are located at two different sites,
then both diquark operators can be chosen to be scalar,
however, they must be antisymmetric in space,  
i.e., with odd integer orbital angular momentum. 
Thus the parity of lowest lying state of
this scalar-scalar-antifermion system is even, as suggested in the
original Jaffe-Wilzcek model. (Note that all correlated quark models
e.g., Karliner-Lipkin model\cite{Karliner:2003dt} and
      flavor-spin model\cite{Stancu:2003if}, 
advocate that the parity of $ \Theta^+(1540) $ is positive.)



\section{Computation of quark propagators}
 
Now it is straightforward to work out the pentaquark propagator
$ \langle \Theta_{x\alpha} \bar\Theta_{y\delta} \rangle $ 
in terms of quark propagators. 
In lattice QCD with exact chiral symmetry,
quark propagator\cite{Chiu:1998eu} with bare mass $ m_q $ 
is of the form $ (D_c + m_q )^{-1} $ 
where $ D_c $ is exactly chirally symmetric
at finite lattice spacing. 
In the continuum limit, $ (D_c + m_q)^{-1} $ reproduces
$ [ \gamma_\mu ( \partial_\mu + i A_\mu ) + m_q ]^{-1} $. 
For the optimal domain-wall fermion\cite{Chiu:2002ir} 
with $ N_s + 2 $ sites in the fifth dimension,   
\BAN
D_c &=& 2m_0 \frac{1 + \gamma_5 S(H_w)}{1 - \gamma_5 S(H_w)}, \hspace{4mm}
S(H_w) = \frac{1 - \prod_{s=1}^{N_s} T_s}
                {1 + \prod_{s=1}^{N_s} T_s}, \\
T_s &=& \frac{1 - \omega_s H_w }{1 + \omega_s H_w},  \hspace{4mm}
H_w = \gamma_5 D_w,  
\EAN
where $ D_w $ is the standard Wilson Dirac operator plus a negative 
parameter $ -m_0 $ ($0<m_0<2$), and $ \{ \omega_s \} $ are a set of 
weights specified by an exact formula such that $ D_c $ 
possesses the optimal chiral symmetry\cite{Chiu:2002ir}. Since 
\BAN
( D_c + m_q )^{-1} 
= (1-rm_q)^{-1} [ D^{-1}(m_q) - r ],  \hspace{4mm} r = \frac{1}{2m_0}
\EAN
and $ D(m_q) = m_q + (m_0 - m_q/2)[ 1 + \gamma_5 S(H_w) ] $,
thus the quark propagator can be obtained by solving 
the system $ D(m_q) Y = \Id $ with nested conjugate 
gradient\cite{Neuberger:1998my},
which turns out to be highly efficient (in terms of the precision 
of chirality versus CPU time and memory storage) if the inner 
conjugate gradient loop is iterated with Neuberger's double pass 
algorithm\cite{Neuberger:1998jk}.

We generate 100 gauge configurations with Wilson gauge action
at $ \beta = 6.1 $ on the $ 20^3 \times 40 $ lattice.
Then we compute two sets of (point-to-point) quark propagators,  
for periodic and antiperiodic boundary conditions in the time 
direction respectively. Here the boundary condition
in any spatial direction is always periodic. 
Now we use the averaged quark propagator to compute the 
time correlation function of any hadronic observable such that 
the effects due to finite $ T $ can be largely
reduced\cite{Chiu:2004xx}.   

Fixing $ m_0 = 1.3 $, we project out 16 low-lying eigenmodes of 
$ |H_w| $ and perform the nested conjugate gradient in the complement
of the vector space spanned by these eigenmodes. For  
$ N_s = 128 $, 
the weights $ \{ \omega_s \} $ are fixed with $ \lambda_{min} = 0.18 $ 
and $ \lambda_{max} = 6.3 $, 
where $ \lambda_{min} \le \lambda(|H_w|) \le \lambda_{max} $
for all gauge configurations.    
For each configuration, (point to point) quark propagators are computed
for 30 bare quark masses in the range $ 0.03 \le m_q a \le 0.8 $, 
with stopping criteria $ 10^{-11} $ and $ 2 \times 10^{-12} $
for the outer and inner conjugate gradient loops respectively.

Then the norm of the residual vector of each column of the 
quark propagator is less than $ 2 \times 10^{-11} $,   
$
|| (D_c + m_q ) Y - \Id || < 2 \times 10^{-11},  
$
and the chiral symmetry breaking due to finite $ N_s (=128) $ is 
less than $ 10^{-14} $,
$
\sigma = \left| Y^{\dagger} S^2 Y/Y^{\dagger} Y - 1 \right|
< 10^{-14},
$
for every iteration of the nested conjugate gradient. 

\section{Determination of $ a^{-1} $ and $ m_s $}

After the quark propagators have been computed, we first 
measure the pion propagator and its time correlation function, and 
extract the pion mass ($ m_\pi a $) 
and the pion decay constant ($ f_\pi a $).
With the experimental input $ f_\pi = 132 $ MeV,  
we determine $ a^{-1} = 2.237(76) $ GeV. 

The bare mass of strange quark is determined by extracting the 
mass of vector meson from the time correlation function
\BAN
C_V (t) = \frac{1}{3} \sum_{\mu=1}^3 \sum_{\vec{x}}
\tr\{ \gamma_\mu (D_c + m_q)^{-1}_{x,0} \gamma_\mu
     (D_c + m_q)^{-1}_{0,x} \}
\EAN
At $ m_q a = 0.08 $, $ M_V a = 0.4601(44) $,
which gives $ M_V = 1029(10) $ MeV, in good agreement with
the mass of $ \phi(1020) $. Thus we take the strange quark 
bare mass to be $ m_s a = 0.08 $. 
Then we have 10 quark masses smaller than $ m_s $, i.e.,  
$ m_u a = 0.03, 0.035, 0.04, 0.045, 0.05, 0.055, 0.06, 0.065, 0.07, 0.075 $. 
In this paper, we work in the isospin limit $ m_u = m_d $. 

\section{The $ 3 \times 3 $ correlation matrix for $ \Theta $}

Next we compute the propagators   
$ \langle (O_i)_{x\alpha} (\bar O_j)_{y\delta} \rangle $ 
with fixed $ y = (\vec{0},0) $, 
and their time correlation functions $ C_{ij}^{\pm}(t) $ with 
$ \pm $ parity
\BAN
C_{ij}^{\pm}(t) = \left< \sum_{\vec{x}} \tr \left[ \frac{1\pm\gamma_4}{2}
\langle O_i(\vec{x},t) {\bar O}_j(\vec{0},0) \rangle_f \right] \right>_U
\EAN
where the trace sums over the Dirac space,   
and the subscripts $ f $ and $ U $ 
denote fermionic average and gauge field ensemble average respectively. 
Then the $ 3 \times 3 $ correlation matrix 
$ C^{\pm}(t) = \{ C_{ij}^{\pm}(t) \} $ can be constructed. 
Now with a judiciously chosen $ t_0 $, we diagonalize the normalized 
correlation matrix $ C^{\pm}(t_0)^{-1/2} C^{\pm}(t) C^{\pm}(t_0)^{-1/2} $
and obtain eigenvalues $ \{ A_i^{\pm}(t) \} $, 
and from which to extract the masses $ \{ m_i^{\pm} \} $ of 
the lowest lying and two excited states for $ \pm $ parity respectively. 
This is the variational method\cite{Michael:1985ne,Luscher:1990ck}
to disentangle the lowest and the excited states. 
Then the mass $ m_i^{\pm} $ can be 
extracted by single exponential fit to $ A_i^{\pm}(t) $,  
for the range of $ t $ in which the effective mass 
$ M_{\mbox{eff}}(t) = \ln [A(t)/A(t+1)]  $
attains a plateau.
 
\begin{figure}[htb]
\begin{center}
\begin{tabular}{@{}cc@{}}
\includegraphics*[height=5.5cm,width=5.0cm]{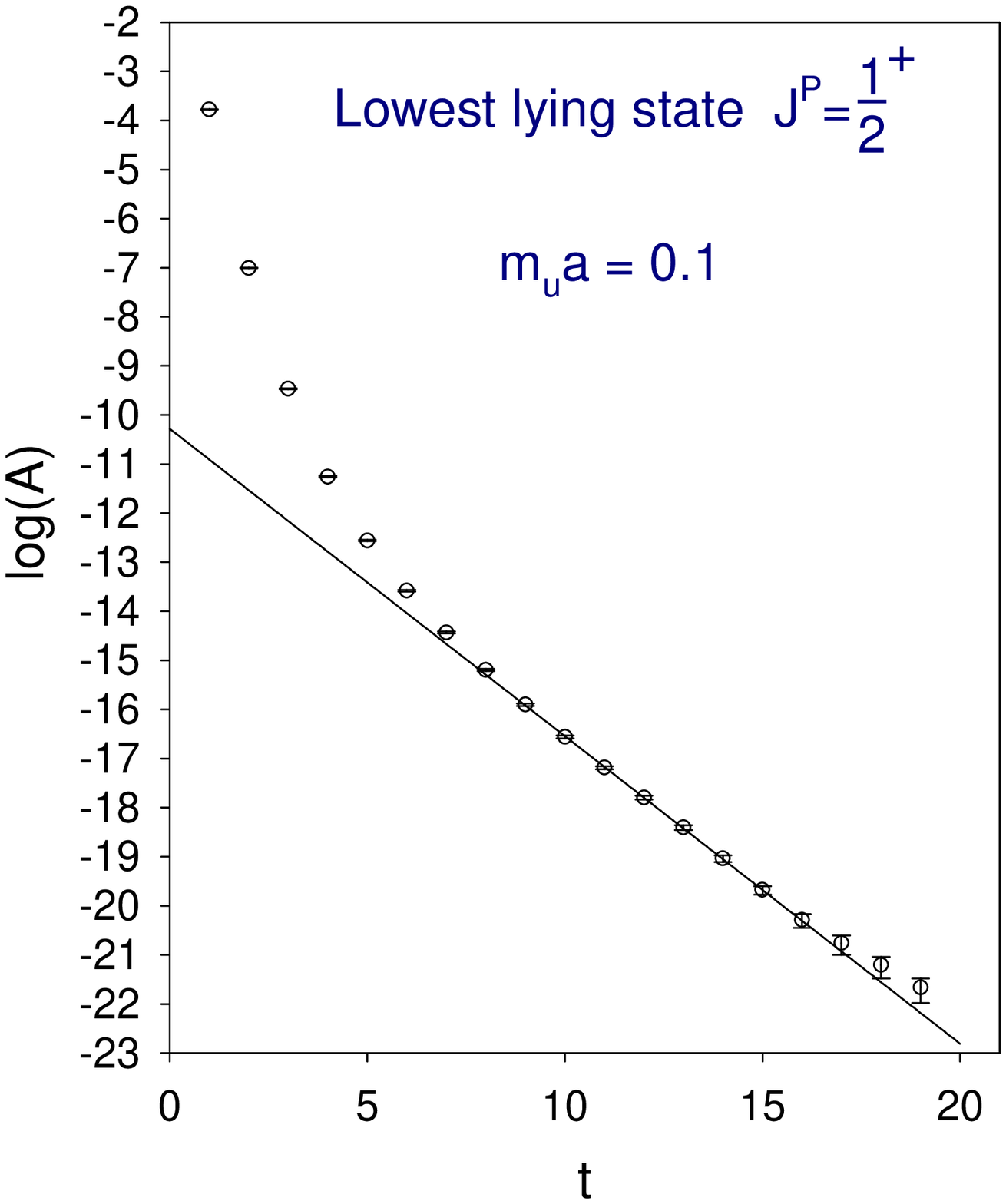}
& 
\includegraphics*[height=5.5cm,width=5.0cm]{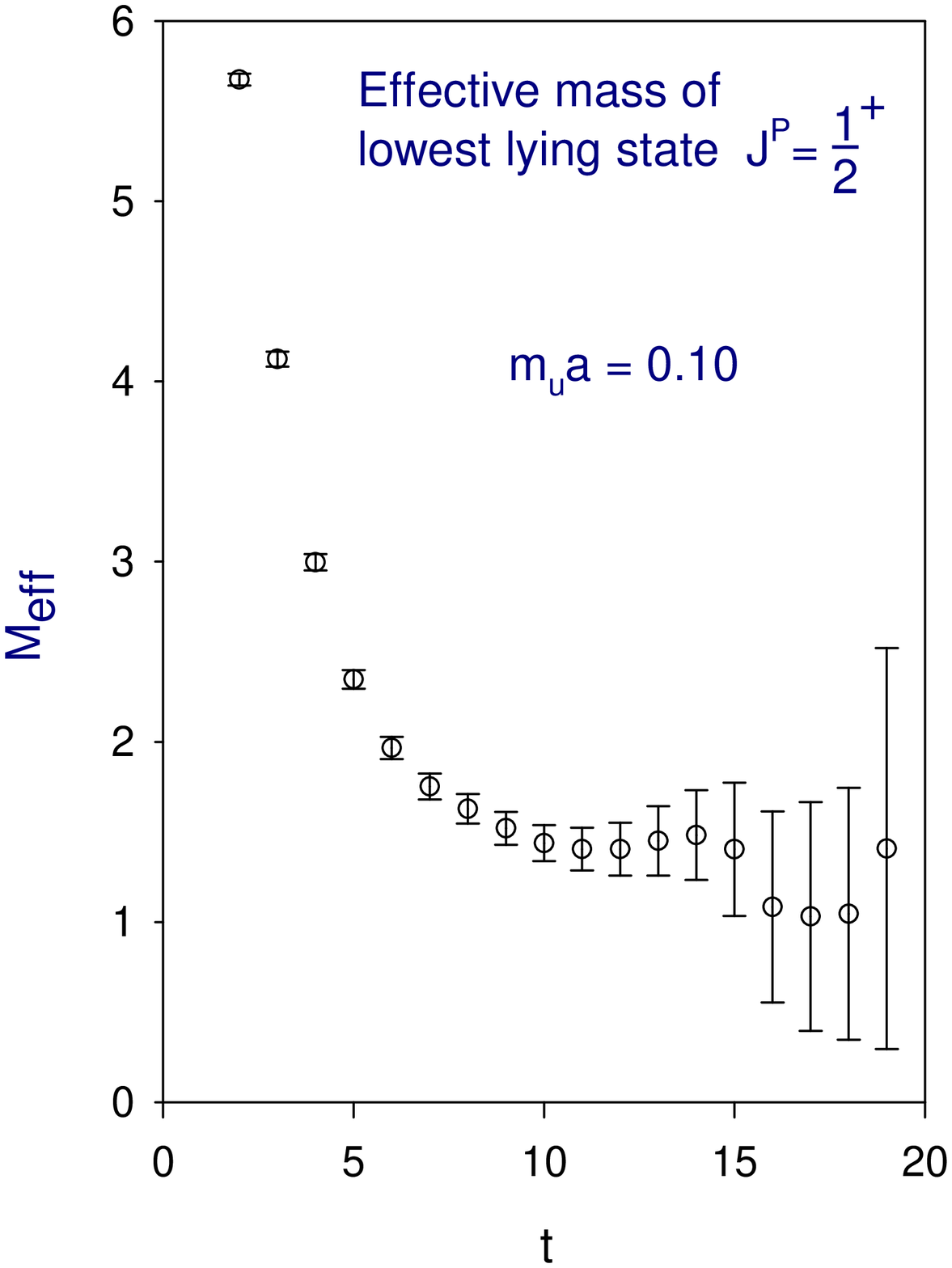}
\\ (a) & (b)
\end{tabular}
\caption{(a) The eigenvalue $ A^+(t) $ of the lowest positive parity state, 
for $ m_u a = 0.1 $.
The solid line is the single exponential fit for $ 9 \le t \le 14 $.
(b) The effective mass $ M_{\mbox{eff}}(t) = \ln [A(t)/A(t+1)]  $
of $ A^+(t) $ in Fig.\ \ref{fig:APe3m10}a.
}
\label{fig:APe3m10}
\end{center}
\end{figure}

\begin{figure}[htb]
\begin{center}
\begin{tabular}{@{}cc@{}}
\includegraphics*[height=5.5cm,width=5.0cm]{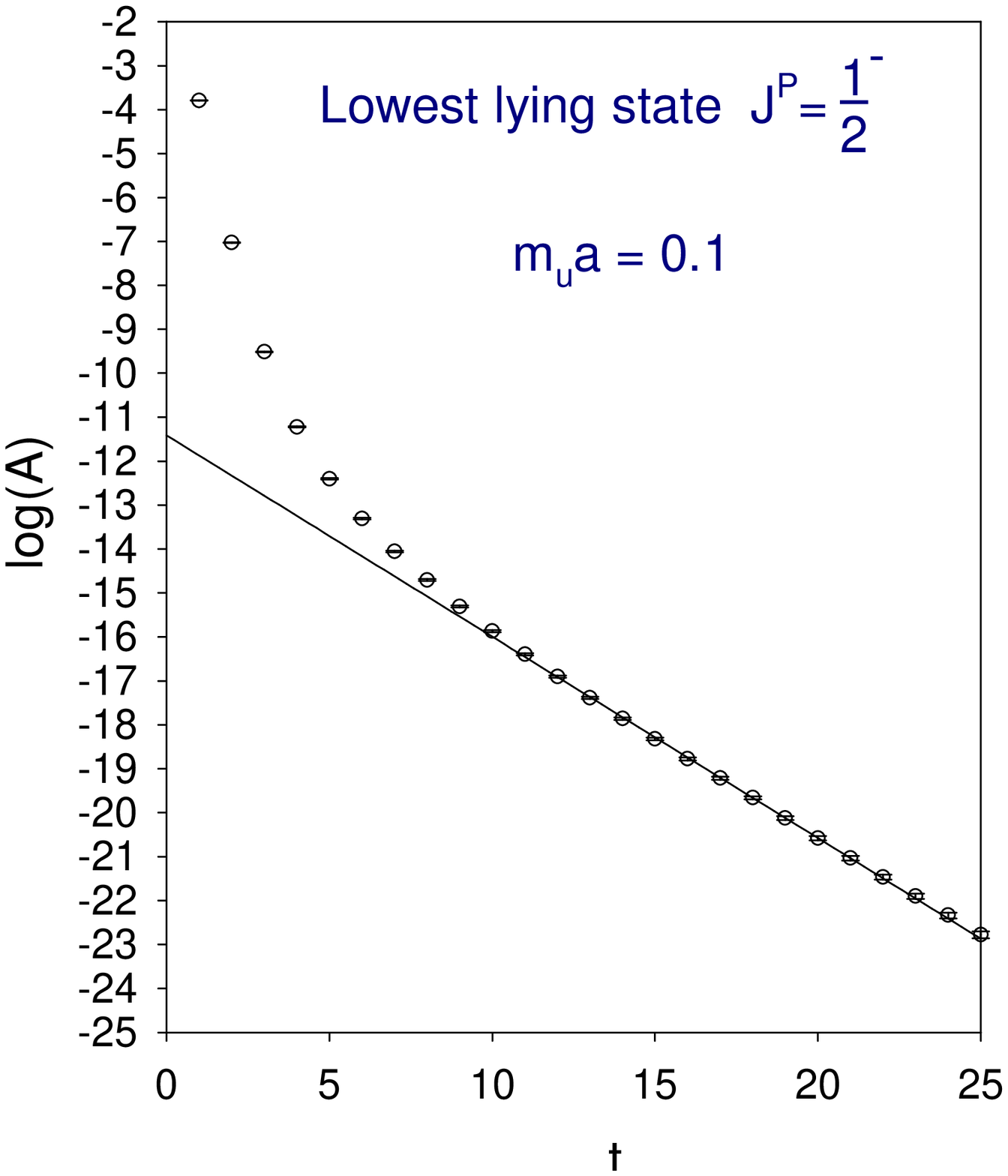}
&
\includegraphics*[height=5.5cm,width=5.0cm]{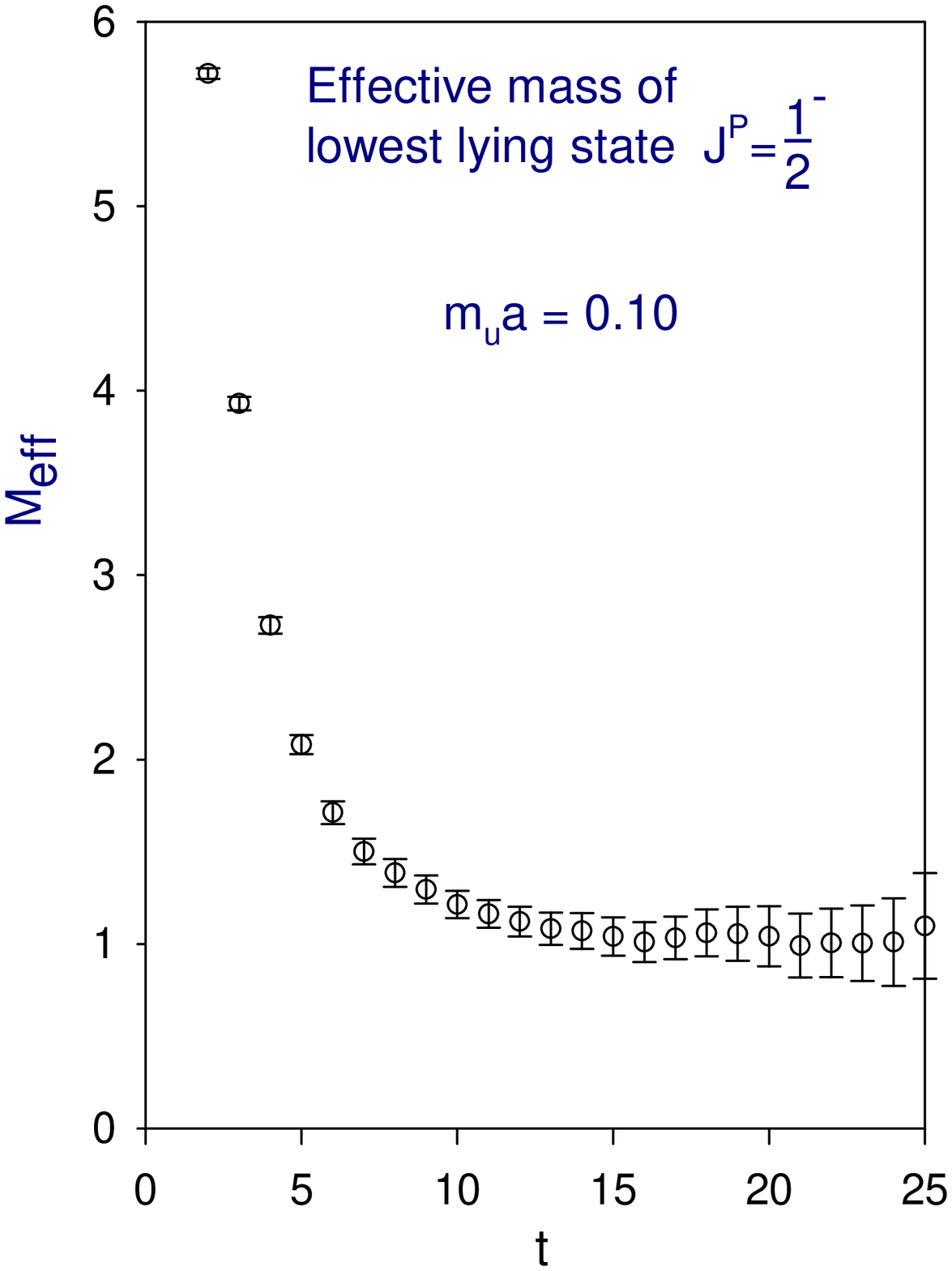}
\\ (a) & (b)
\end{tabular}
\caption{(a) The eigenvalue $ A^-(t) $ of the lowest 
negative parity state, for $ m_u a = 0.1 $.
The solid line is the single exponential fit for $ 11 \le t \le 17 $.
(b) The effective mass of $ A^{-}(t) $ in Fig.\ \ref{fig:AMe1m10}a.}
\label{fig:AMe1m10}
\end{center}
\end{figure}

In Fig.\ \ref{fig:APe3m10}a, the eigenvalue $ A^+(t) $ corresponding to 
the lowest $ J^P = 1/2^+ $ state is plotted versus the time slices, 
for $ m_u a = 0.1 $, while the corresponding effective mass is shown 
in Fig.\ \ref{fig:APe3m10}b. 
Here we have suppressed any data point which has error 
(jackknife with single elimination) larger than 
its mean value. Similarly, the eigenvalue $ A^{-}(t) $
corresponding to the lowest $ J^P=1/2^- $ state and 
its effective mass are plotted in Fig.\ \ref{fig:AMe1m10}a
and Fig.\ \ref{fig:AMe1m10}b. 

First, we observe that the effective mass of the lowest $ J^P = 1/2^+ $ 
state in 
Fig.\ \ref{fig:APe3m10}b 
attains a plateau for $ t \in [9,14] $, and its mass can be extracted by 
single exponential fit (Fig.\ \ref{fig:APe3m10}a).
Similarly, the effective mass of the lowest $ J^P = 1/2^- $ state in
Fig.\ \ref{fig:AMe1m10}b
attains a plateau for $ t \in [11,17] $, and its mass can be extracted by 
single exponential fit (Fig.\ \ref{fig:AMe1m10}a).

In Fig.\ \ref{fig:mtheta}a,  
the masses of the lowest lying $ J=(1/2)^\pm $ states are plotted 
versus $ m_\pi^2 $, where all mass fits have $ \chi^2/d.o.f. < 1 $. 
Using the four smallest masses (i.e., with 
$ m_u a = 0.03, 0.035, 0.04, 0.045 $) 
for chiral extrapolation (linear in $ m_\pi^2 $ ) 
to physical pion mass $ m_\pi = 135 $ MeV, 
we obtain the mass of the lowest lying states: 
$ m(1/2^-) = 1424 (57) $ MeV, and $ m(1/2^+) = 1562(121) $ MeV, 
in agreement with the masses of  
$ m_K + m_N \simeq 1430 $ MeV, and $ \Theta(1540) $ respectively. 

\section{Distinguishing the KN scattering states}

Now the question is whether they are scattering states
or bound states. In order to obtain the mass spectrum of $ KN $ 
scattering states, we consider the time correlation function 
of $ KN $ operator without any exchange of quarks between 
$ K $ and $ N $ in its propagator, i.e., the interaction 
between $ K $ and $ N $ is only through the exchange of gluons. 
Explicitly, 
\BAN
C_{KN}(t) = \left< \sum_{\vec{x}} 
                    \langle N(\vec{x},t) \bar N(\vec{0},0) \rangle_f
            \langle K(\vec{x},t) \bar K(\vec{0},0) \rangle_f \right>_U
\EAN
where $ N=[{\bf u}^T C \gamma_5 {\bf d}] {\bf d} $, and 
$ K=\bar{\bf s}\gamma_5{\bf u} $. 

\begin{figure}[htb]
\begin{center}
\begin{tabular}{@{}cc@{}}
\includegraphics*[height=5.5cm,width=5.0cm]{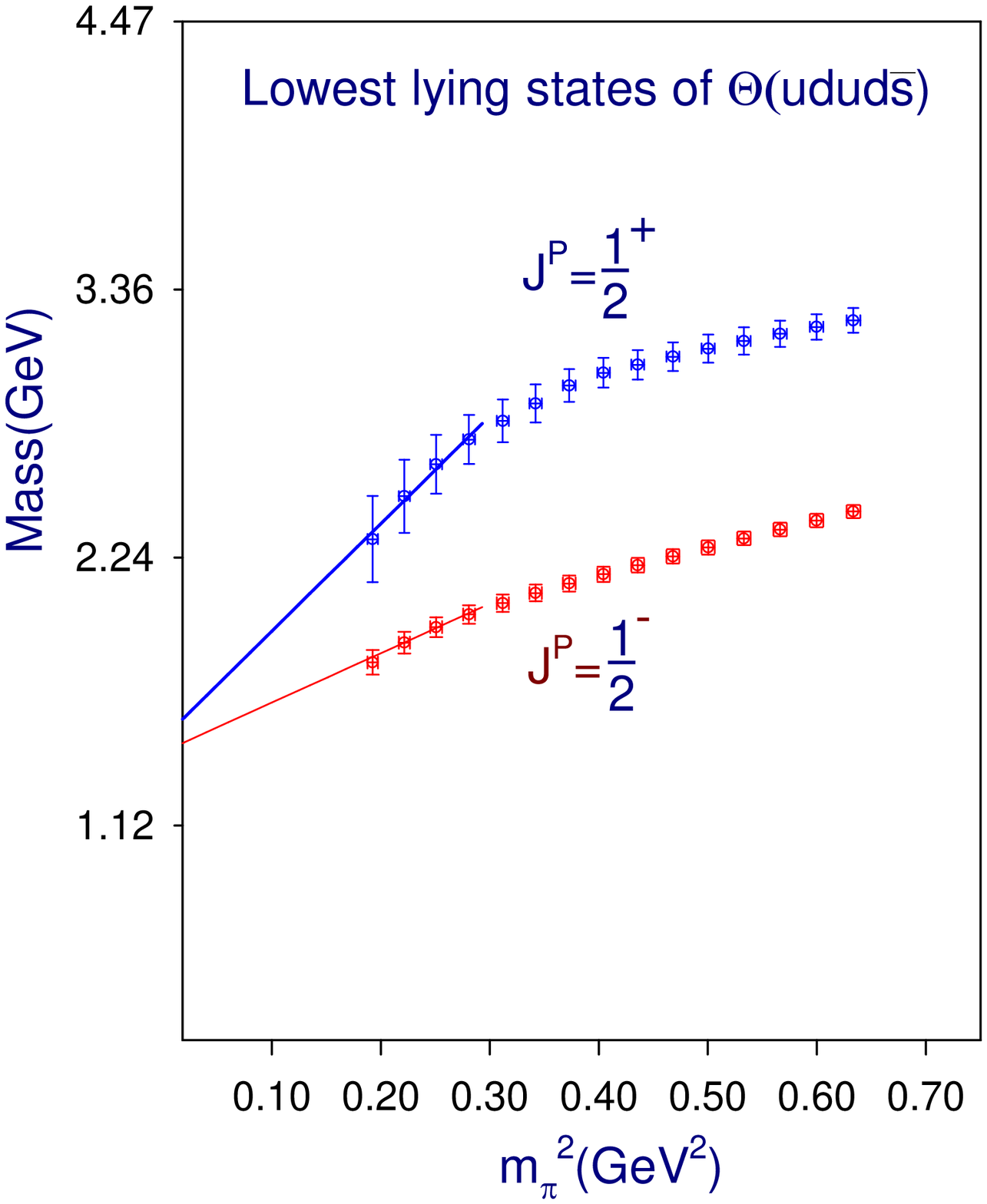}
&
\includegraphics*[height=5.5cm,width=5.0cm]{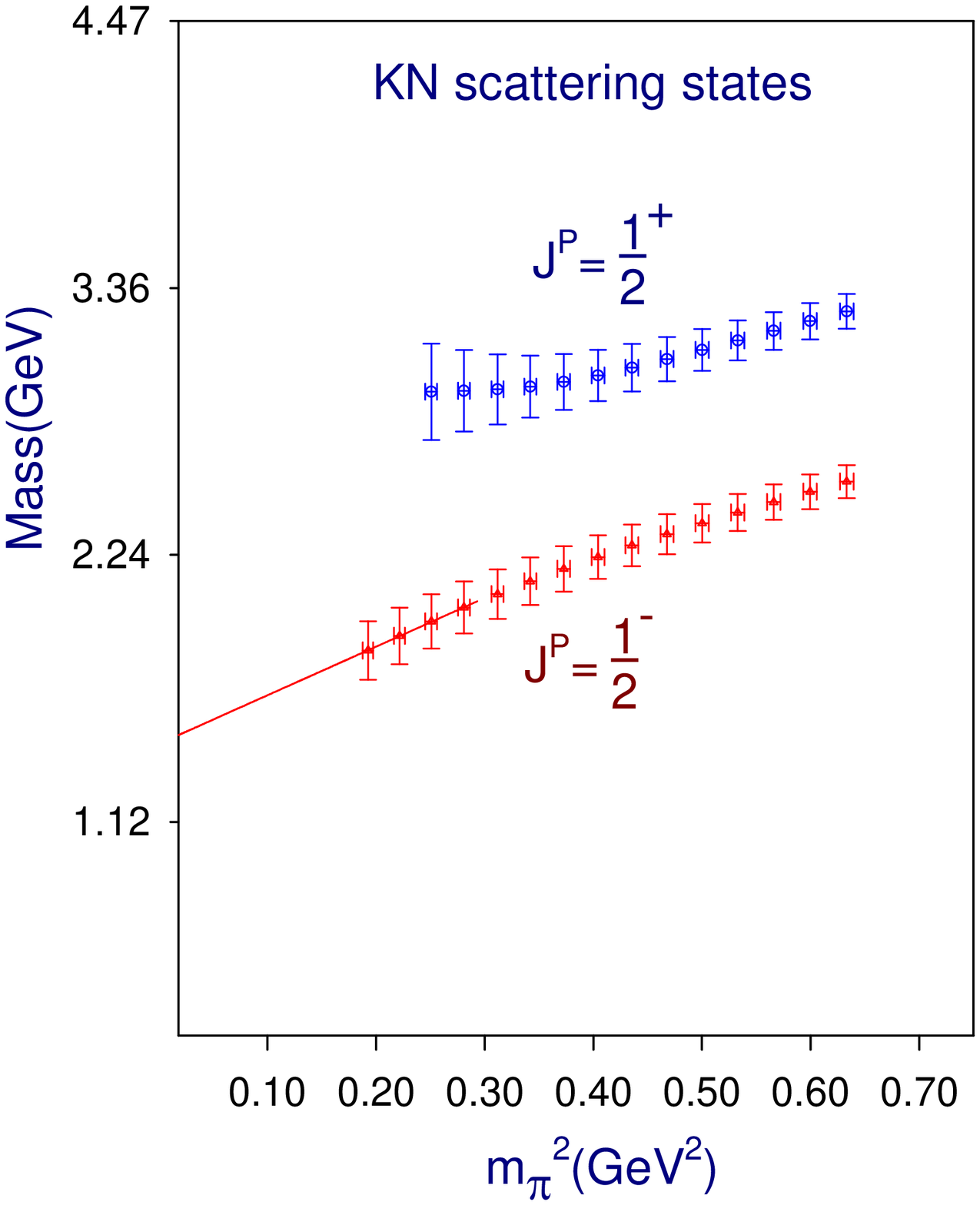}
\\ (a) & (b)
\end{tabular}
\caption{(a) The masses of the lowest lying states of $ \Theta(udud \bar s) $.
The solid lines are chiral extrapolation (linear in $ m_\pi^2 $) 
using four smallest masses.
(b) The masses of the lowest lying $ KN $ scattering states. 
The solid line (for $ J^P = 1/2^- $) is the chiral extrapolation  
using four smallest masses.}
\label{fig:mtheta}
\end{center}
\end{figure}

The masses of lowest lying $ KN $ scattering states are plotted in 
Fig. \ref{fig:mtheta}b. For the $ J^P = 1/2^- $ state, 
using the four smallest masses for chiral extrapolation  
to $ m_\pi = 135 $ MeV, we obtain $ m_{KN}(1/2^-) = 1433(72) $ MeV, 
in agreement with the mass of $ m_K + m_N \simeq 1430 $ MeV, 
the KN s-wave. Further, its mass spectrum is almost identical 
to that of the lowest $ J^P = 1/2^- $ state of $ \Theta $ 
in Fig. \ref{fig:mtheta}a, for the entire range of $ m_u $. 
Thus we identify the lowest $ J^P = 1/2^- $ state of $ \Theta(udud\bar s) $ 
with the KN s-wave scattering state. 

On the other hand, for the $ J^P = 1/2^+ $ state, 
its mass is much higher than the naive estimate 
$ \sqrt{m_K^2 + (2\pi/L)^2} + \sqrt{m_N^2 + (2\pi/L)^2} $, 
where $ L $ is the lattice size in spatial directions. 
This suggests that the KN p-wave (in the quenched approximation) 
in a finite torus is much more complicated than two free particles
with momenta $ \vec{p}_K = -\vec{p}_N = 2\pi\hat{e_i}/L $.
Further, the mass of KN p-wave scattering state in Fig. \ref{fig:mtheta}b 
is always larger than the mass of the $ J^P = 1/2^+ $ state 
in Fig. \ref{fig:mtheta}a. In particular, for $ m_u < m_s $, 
the former is significantly larger than the latter. 
This seems to suggest that the lowest $ J^P = 1/2^+ $ state of 
$ \Theta(udud \bar s) $ is {\it different} from the KN p-wave scattering state. 
In other words, it is likely to be a bound state with mass $ 1562(121) $ MeV. 
If it is identified with $ \Theta^+(1540) $, then it
predicts that {\it the parity of $ \Theta^+(1540) $ is positive}.

\section{Concluding remarks}

It is vital to re-confirm the above picture with a larger lattice, smaller 
lattice spacing, and higher statistics, especially for the regime 
$ m_u < m_s $, which is crucial for the chiral extrapolation. 
Further, to ensure that the lowest lying $ J^P = 1/2^+ $ is 
{\it not} a scattering state, one can also compute its 
spectral weight for two different volumes\cite{Mathur:2004jr},
since for a scattering state, its spectral weight 
is inversely proportional to the volume. 
Even if it is confirmed to be a bound state, one still 
has to find out whether its decay width 
could be as small as $ 20 $ MeV, compatible to that of 
$ \Theta^+(1540) $. Our present data only shows an unambiguous 
signal of pentaquark $ (udud \bar s) $ resonance around 1540 MeV, 
with $ S = +1 $ and $ I(J^P) = 0(1/2^+) $.


%

\end{document}